\documentclass[epj]{svjour}
\usepackage[latin1]{inputenc}
\usepackage[dvips]{graphicx}
\usepackage{amsmath}
\usepackage{bm}
\usepackage{amssymb}
\usepackage{latexsym}
\usepackage{dcolumn}
\usepackage{url}
\usepackage{color}
\newcolumntype{d}{D{.}{.}{-1}}
\newcolumntype{f}[1]{D{.}{.}{#1}}
\def\etal{\textit{et al. }}
\def\asc#1{ {\color{blue} {... #1 ...} } }               
\def\asc2#1{{\color{blue}{#1}}}  
\def\asq#1{ {\color{red} {..(??. #1 .??)..} } }         
\def\asq2#1{ {\color{red} {? #1 ?} } }         

\begin{document}
%
%
%
   \title{Relativistic transition wavelenghts and probabilities for spectral lines of Ne II}
%

%
%
%
%
%
\author{J.\ P.\ Santos\inst{1}\thanks{jps@fct.unl.pt} \and
        A.\ M.\ Costa\inst{2} \and C.\ Madruga\inst{1}  
        \and F.\ Parente\inst{1} \and P.\ Indelicato\inst{3} 
       }
%
%
\offprints{J. P. Santos}   
\institute{Centro de F{\'\i}sica At{\'o}mica, Departamento de
  F{\'\i}sica, Faculdade de Ci{\^e}ncias e 
  Tecnologia, FCT,  Universidade Nova de Lisboa, Monte de Caparica,
  2829-516 Caparica, Portugal 
  \and Centro de F{\'\i}sica At{\'o}mica, Departamento de F{\'\i}sica,
  Faculdade de Ci{\^e}ncias, FCUL, Universidade de Lisboa, Campo
  Grande, Ed. C8, 1749-016 Lisboa, Portugal 
  \and Laboratoire Kastler Brossel, \'Ecole Normale Sup\'erieure,
  CNRS, Universit\'e P. et M. Curie -- Paris 6, Case 74; 4, place
  Jussieu, 75252 Paris CEDEX 05, France 
  }

%
%
\date{Received: \today / Revised version: date }
\date{\today}
%
%
%
%

 \abstract {Transition wavelengths and probabilities for several  2p$^4$ 3p - 2p$^4$ 3s and 2p$^4$ 3d - 2p$^4$ 3p lines in  fluorine-like neon ion (NeII) have been calculated within the  multiconfiguration Dirac-Fock (MCDF) method with quantum  electrodynamics (QED) corrections. The results are compared with  all existing experimental and theoretical data.}

%
%

\PACS{31.15.vj Electron correlation calculations for atoms and ions:
  excited states, 32.70.Cs Oscillator strengths, lifetimes,
  transition moments, 32.30.Jc Visible and ultraviolet spectra  }

   \maketitle
%

\section{Introduction}
\label{sec1}

Knowledge of accurate atomic parameters, such as transition
probabilities, is fundamental in the study of atomic structure,
and of laboratory or astrophysical plasmas. In what concerns
astrophysics, these parameters are crucial to estimate the densities
of species in the atmospheres of stars, galaxies and
nebulae~\cite{3258}.

Neon, after H, He, O, and C, is one of the most abundant elements in
Universe and is one of the products of hydrogen and helium
thermonuclear reactions in the orderly evolution of stellar
interiors~\cite{3259}.

Because of its cosmic abundance and atomic properties all neon ions
are of importance in various astrophysical sources, in particular Ne II,
whose emission lines are very intense.

The interest in the spectrum of Ne II has been further stimulated by
large discrepancies between the existing transition probability values
of the spontaneous emission rates (Einstein's $A_{ki}$ values) for weak
intercombination, or spin-forbidden, lines.

%
%


Koopman~\cite{1029} performed the first extensive measurements of Ne II spectra, covering around 50 lines, using an electrically driven shock tube as the spectroscopic source and adjusted in some cases with the help of the $J$-file sum rule proposed by Condon an Shortley~\cite{133}. This rule, used in systems that do not accurately follow the LS coupling scheme, relates the sum of the oscillator strengths of the lines with a common upper or lower level to the quantum weight of that state.
%
Koopman's work is one of the original references given in the NIST-ASD compilation~\cite{853}.



Extensive spontaneous-emission data on Ne II using a water-cooled
neon-ion-laser discharge without any mirrors as the spectroscopic
source were obtained by \cite{1115}.


After more than twenty years since the last reported measured values,
Burshtein and Vujnovic~\cite{1039} obtained the absolute transition probabilities of the
3p-3s transition array within an overall uncertainty of 25\%.

Griesman \etal~\cite{1193}, using a high-current hollow-cathode lamp,
provided absolute measurements of transition probabilities for 48 Ne
II lines, and relative measurements for 83 Ne II lines, corresponding
to 3p, 3d and 4s upper levels and including many weak intersystem
lines for the first time. Their uncertainty is 11\% for strong lines
($A_{ki}>0.1 \times 10^{8}$ s$^{-1}$), increasing to 32\% for weaker
lines.

In 1999, Fuhr and Wiese~\cite{1113} published a critical compilation
of atomic transition probabilities for about 9000 selected lines of
all elements, mainly for neutral and singly ionized spectra, to which
they assigned an uncertainty band of 25-50\%.  

Using a pulsed discharge lamp emission experiment, Val \etal~\cite{1897} reported transition
probability values for 94 Ne II lines, in the 337-463 nm spectral
region. 
Their absolute $A_{ki}$ values were obtained by using bibliographic
data as reference and, consequently, their quality is linked to the
intrinsic quality of these data. They estimated an average error
around 15\% for $A_{ki}$ transition values larger than $0.2 \times
10^{8}$ s$^{-1}$ and around 30\% for the weakest transitions.

Djenize \etal~\cite{2189} measured the transition probabilities of 42
Ne II spectral lines in a linear low-pressure pulsed arc. The
transition probabilities were obtained using the relative line
intensity ratio method. Their uncertainties range from 10\% to 20\%.

Using a high-current hollow cathode discharge in pure neon
Bridges and Wiese~\cite{2122} studied experimentally weak intersystem lines and related
strong persistent lines of Ne II. They obtained transition
probabilities for some 3p-3s, 3d-3p and 4f-3d lines with uncertainties
smaller than 25\%, 27\% and 31\%, respectively.


%
%

In what concerns the theoretical work, Garstang ~\cite{1227}
performed, to our knowledge, the first Ne II ion transition
probability calculations using the intermediate coupling approximation
and parameters determined empirically from observational data.

Marantz~\cite{M007} calculated, within the intermediate coupling
approximation, the radial wavefunctions by using the computer program
developed by Herman and Skillman~\cite{3351}, and obtained results
very similar to those given by Garstang.

Using the general configuration interaction code CIV3, Blackford and Hibbert~\cite{1200}
reported calculations for transitions in F-like ions in the range
$Z=10, \ldots, 33$. Only a limited amount of correlation was included
within the $n=2$ electrons, and the Breit-Pauli Hamiltonian included
only the mass correction, Darwin terms, and a one-electron spin-orbit
operator.  These authors have made some adjustments to the transition
energies to bring them close to experimental energy separations.


Froese Fischer and He~\cite{1027} performed calculations for some Ne
II transitions taking two approaches: the multiconfiguration
Hartree-Fock (MCHF) method with Breit-Pauli corrections, omitting only
the orbit-orbit term, which does not contribute to term mixing, and
the multiconfiguration Dirac-Hartree-Fock (MCDHF) method with Breit
correction. The latter method allowed the authors to compute the
results in two different gauges: in the Coulomb (velocity) gauge and
Babushkin (length) gauge.

Godefroid and Hibbert~\cite{1036} re-analyzed some CIV3 results for Ne
II and found that the major reason for the discrepancy with the
earlier MCHF data by Froese Fischer and He~\cite{1027} was mainly
related to the omission of the ``fine-tuning'' in the previous
calculations, which is made by correcting empirically the off-diagonal
coupling Hamiltonian matrix element, assuming a proportionality
between the coupling term and the fine-structure energy separation.

Zheng and Wang~\cite{1026} employed the Weakest Bound Electron Potential Model
(WBEPM) method to calculate some transition probabilities for
individual lines of Ne II. The needed parameters for this
method were determined by fitting the experimental value of
energy level and the expectation value of radial distance.

Recently, Froese Fischer and Tachiev~\cite{1107} reported energy
levels, lifetimes, and transition probabilities for, among others, the
F-like ($Z= 9, \ldots, 14$) sequence.
The wavefunctions were
determined using the MCHF method with relativistic effects included
through the Breit-Pauli Hamiltonian, omitting only the orbit-orbit
interaction. 
Afterwards, the calculated transition probabilities were ``adjusted" by correcting the transition energy using the available experimental data.
The authors reported that the shifts in the
energy adjustments were considerably smaller than in the earlier MCHF
calculations for Ne II~\cite{1027} due to the inclusion of more
correlation, and concluded that the MCDHF levels published in
reference~\cite{1027} were not correctly ordered.

In the present \textit{ab initio} theoretical work we start from a
Dirac-Fock calculation with Breit interaction included
self-consistently.  Higher-order retardation and one-electron
radiative corrections are also included, and the screening of the
self-energy is evaluated using the Welton approximation.  Correlation
is added within the multiconfiguration Dirac-Fock method (MCDF). In
this framework we have calculated the relativistic transition
wavelengths for several Ne II 2p$^4$ 3p - 2p$^4$ 3s and 2p$^4$ 3d -
2p$^4$ 3p lines, and used them to compute the transition
probabilities.

%
%
\section{Relativistic calculations}

Ne II has seven electrons outside the 1s$^{2}$ core, which produce
interactions in the excited states. In order to obtain accurate
results in an \textit{ab initio} based in the Configuration
Interaction (CI) method, a large number of configurations should be
considered to deal with these interactions, making the calculation
very complex and time consuming. Another problem of this \textit{ab
  initio} method is to select a proper configuration wave function,
especially for excited states, because different selections of
configuration wave function make accuracy of different results.

The MCDF approach, being a variational method, has the advantage of
providing, with smaller basis sets, results of the same accuracy of
those obtained with the CI method, or variants.

The general relativistic program developed by Desclaux and Indelicato
~\cite{32,62,3226} was used to compute the energies and wavefuntions,
as well as radiative transition probabilities within the MCDF method.
%

In order to obtain a correct relationship between many-body methods
and quantum electrodynamics (QED), one should start from the no-pair
Hamiltonian~\cite{47,231,229,230}
\begin{equation}
\label{eq:hamilnopai}
        {\cal H}^{\mbox{{\tiny no pair}}}=\sum_{i=1}^{N}{\cal
        H}_{D}(r_{i})+\sum_{i<j}{\cal V}(|\bm{r}_{i}-\bm{r}_{j}|),
\end{equation}
where ${\cal H}_{D}$ is the one electron Dirac operator and ${\cal V}$
is an operator representing the electron-electron interaction of order
one in $\alpha$, properly set up between projection operators
$\mathit{\Lambda}_{ij}^{++}=\mathit{\Lambda}_{i}^{+}\mathit{\Lambda}_{j}^{+}$
to avoid coupling positive and negative energy states
\begin{equation}
\mathcal{V}_{ij}=\mathit{\Lambda}_{ij}^{++}V_{ij}\mathit{\Lambda}_{ij}^{++}.
\end{equation}
The expression of $V_{ij}$ in the Coulomb gauge and in atomic units is
\begin{eqnarray}
\label{eq:coulop}
V_{ij} =& \,\,\,\, \frac{1}{r_{ij}}  \\
\label{eq:magop}
&-\frac{\boldsymbol{\alpha}_{i} \cdot \boldsymbol{\alpha}_{j}}{r_{ij}} \\ 
\label{eq:allbreit}
& - \frac{\boldsymbol{\alpha}_{i} \cdot
         \boldsymbol{\alpha}_{j}}{r_{ij}} 
[\cos\left(\frac{\omega_{ij}r_{ij}}{c}\right)-1]
         \nonumber \\
& + c^2(\boldsymbol{\alpha}_{i} \cdot
         \boldsymbol{\nabla}_{i}) (\boldsymbol{\alpha}_{j} \cdot
         \boldsymbol{\nabla}_{j})
         \frac{\cos\left(\frac{\omega_{ij}r_{ij}}{c}\right)-1}{\omega_{ij}^{2} r_{ij}},
\end{eqnarray}
%
%
where $r_{ij}=\left|\bm{r}_{i}-\bm{r}_{j}\right|$ is the
inter-electronic distance, $\omega_{ij}$ is the energy of the
exchanged photon between the two electrons, $\bm{\alpha}_{i}$ are the
Dirac matrices and $c$ is the speed of light~\cite{63}.

The term (\ref{eq:coulop}) represents the Coulomb interaction, the
term (\ref{eq:magop}) is the Gaunt (magnetic) interaction, and the
last two terms (\ref{eq:allbreit}) stand for the retardation operator.
In this expression the $\bm{\nabla}$ operators act only on $r_{ij}$
and not on the following wave functions.

By a series expansion of the operators in expressions~(\ref{eq:magop})
and (\ref{eq:allbreit}) in powers of $\omega_{ij}r_{ij}/c \ll 1$ one
obtains the Breit interaction, which includes the leading retardation
contribution of order $1/c^{2}$. The Breit interaction is, then, the
sum of the Gaunt interaction (\ref{eq:magop}) and the Breit
retardation
\begin{equation}
\label{eq:breit}
B^{\textrm{\scriptsize{R}}}_{ij} =
{\frac{\bm{\alpha}_i\cdot\bm{\alpha}_j}{2r_{ij}}} - 
\frac{\left(\bm{\alpha}_i\cdot\bm{r}_{ij}\right)\left(\bm{\alpha}_j
\cdot\bm{r}_{ij}\right)}{{2r_{ij}^3}}.
\end{equation}
In the many-body part of the calculation the electron-electron
interaction is described by the sum of the Coulomb and the Breit
interactions. Higher orders in $1/c$, deriving from the difference
between Eqs.~(\ref{eq:allbreit}) and (\ref{eq:breit}) are treated here
only as a first order perturbation.

%
All calculations are done for a finite nucleus using a uniformly charged
sphere. The atomic mass and the nuclear radius were taken from the
tables by Audi \etal~\cite{3275} and Angeli~and \cite{1109}, respectively.

Radiative corrections are also introduced, from a full QED treatment.
The one-electron self-energy is evaluated using the one-electron
values of Mohr and co-workers ~\cite{114,116,3141,PI_01} and corrected for
finite nuclear size~\cite{117}.

The self-energy screening and vacuum polarization are treated with an
approximate method developed by Indelicato and
co-workers~\cite{58,56,53,847,PI_02}. 

For each transition, initial and final states are computed
independently to get accurate correlation energies.  Consequently, the
initial and final state orbitals of identical symmetry are not
orthogonal (see, e.g.~\cite{782} and references therein). This
non-orthogonality is properly taken in account in the transition
probability calculation using L\"{o}wdin's method ~\cite{608}. Being a
fully relativistic method, initial and final levels for each
transition are defined in $jj$-coupling.  However, for comparison with
other published work where levels are characterized by their $LSJ$
values, we show in the tables, for each level, the most important
$LSJ$ set of values which results from the expansion of the $jjJ$ wave
function in terms of $LSJ$ ones.

%
%
\section{Results and discussion}
\label{sec3}

When the wave functions are determined variationally, the energy of a
particular atomic level will be the lowest that can be achieved with
the specific form used for the wave function.  Improvements in the
wave function, such as the inclusion of more configuration state
functions to account for electronic correlation, will lead, in
principle, to monotonic reductions in the energy, which guarantees the
improvement of the energy accuracy.

%
%
%
%
%
\begin{table*}									
%
\caption{Correlation effect on  $1\textrm{s}^{2}$ $2\textrm{s}^{2}$ $2\textrm{p}^{4}$  $3\textrm{p}$ $^{4}$D$_{7/2}\rightarrow$  $1\textrm{s}^{2}$ $2\textrm{s}^{2}$ $2\textrm{p}^{4}$  $3\textrm{s}$ $^{4}$P$_{5/2}$  transition wavelength  $\lambda$ (in nm) and transition probability $A_{ki}$ (in 10$^{8}$ s$^{-1}$). 	The $1\textrm{s}^{2}$ $2\textrm{s}^{2}$ core is omitted in the table entries.							
\label{tab_A}}									
%
%
\begin{tabular}{llccc}									
\hline									
\hline									
\\
Upper level ($^{4}$D$_{7/2}$)	&	Lower level ($^{4}$P$_{5/2}$)  	&	 $\lambda$ 	&	  $A_{ki,\mathrm{length}}$  	&	  $A_{ki,\mathrm{velocity}}$  	\\
\hline									
	&		&		&		&		\\
2p$^{4}$ 3p 	&	2p$^{4}$ 3s	&	347.26	&	1.733	&	1.659	\\
\ \ +\ All correlation until 3d	&	\ \ +\ All correlation until 3d	&	319.51	&	2.139	&	1.895	\\
\ \ +\ 2p$^{2}$ 3p 4s$^{2}$ + 2p$^{2}$ 3p 4p$^{2}$ 	&	\ \ +\ 2p$^{2}$ 3s 4s$^{2}$ + 2p$^{2}$ 3s 4p$^{2}$ 	&	332.57	&	1.905	&	1.832	\\
\ \ +\ 2p$^{2}$ 3p 4d$^{2}$ + 2p$^{2}$ 3p 4f$^{2}$ 	&	\ \ +\ 2p$^{2}$ 3s 4d$^{2}$ + 2p$^{2}$ 3s 4f$^{2}$ 	&	333.03	&	1.897	&	1.829	\\
\ \ +\ 2p$^{2}$ 3p 5s$^{2}$ + 2p$^{2}$ 3p 5p$^{2}$ 	&	\ \ +\ 2p$^{2}$ 3s 5s$^{2}$ + 2p$^{2}$ 3s 5p$^{2}$ 	&	333.85	&	1.883	&	1.825	\\
\ \ +\ 2p$^{2}$ 3p 5d$^{2}$ + 2p$^{2}$ 3p 5f$^{2}$ 	&	\ \ +\ 2p$^{2}$ 3s 5d$^{2}$ + 2p$^{2}$ 3s 5f$^{2}$ 	&		&		&		\\
\ \ +\ 2p$^{2}$ 3p 5g$^{2}$ 	&	\ \ +\ 2p$^{2}$ 3s 5g$^{2}$ 	&	333.94	&	1.882	&	1.824	\\
\ \ +\ 2p$^{2}$ 3p 6s$^{2}$ + 2p$^{2}$ 3p 6p$^{2}$ 	&	\ \ +\ 2p$^{2}$ 3s 6s$^{2}$ + 2p$^{2}$ 3s 6p$^{2}$ 	&	334.06	&	1.880	&	1.824	\\
\ \ +\ 2p$^{2}$ 3p 6d$^{2}$ 	&	\ \ +\ 2p$^{2}$ 3s 6d$^{2}$ 	&	334.23	&	1.878	&	1.814	\\
\ \ +\ 2p$^{2}$ 3p 6f$^{2}$ 	&	\ \ +\ 2p$^{2}$ 3s 6f$^{2}$ 	&	334.23	&	1.878	&	1.814	\\
\\                             									
\hline									
Experimental                        	&	& 333.48$^{\S}$  	&	\multicolumn{2}{c}{1.77$\pm$0.15 $^{\mathrm{\dag}}$}					\\
\\									
\hline									
\hline									
\end{tabular}									
\\									
$^{\S}$ Ref.~\cite{1647}.	\\								
$^{\mathrm{\dag}}$ $A_\mathrm{Exp_\mathrm{Av}}$.									
%
%
\end{table*}									
%
%

Unlike in the evaluation of energies, there is no guaranteed monotonic
improvement in the energy separations between atomic levels
(transition energy) with respect to improvements in the wave function.
Nevertheless, this problem can be overcome by optimizing, in a
systematic manner, both the wave function and energy of each level, so
that the two states are treated, as far as possible, in a balanced
way.  Consequently, the amount of correlation included in the
calculation of the wave function and energy must be equivalent for both initial
and final levels.

%
%
%
\begin{table*}
%
%
\caption{Wavelengths $\lambda$ (in nm) and transition probabilities
    $A_{ki}$ (in 10$^{8}$ s$^{-1}$) for some $3p-3s$ lines and
    comparisons with other theoretical and experimental results.
    $\lambda_\mathrm{TW}$, $A_{\mathrm{TW,l}}$ and $A_{\mathrm{TW,v}}$
    represent, respectively, the wavelength and the transition
    probability values in the length and velocity gauges calculated in
    this work.  $A_\mathrm{G}$, $A_\mathrm{M}$, $A_\mathrm{BH}$,
    $A_\mathrm{Z}$ and $A_\mathrm{FT}$ denote theoretical values taken
    from Garstang~\cite{1227}, Marantz~\cite{M007}, Blackford and
    Hibbert~\cite{1200}, Zheng~\cite{1026}, and Fischer and
    Tachiev~\cite{1107}, respectively.  $\lambda_\mathrm{NIST}$
    represent the experimental NIST values by Kramida and
    Nave~\cite{1647}, and $A_\mathrm{Exp_\mathrm{Av}}$ the weighted
    average experimental values.  
\protect\label{tab_01}}
%
\begin{tabular}{ccccccccccccc}																								
\hline																									
\hline																									
	&	Upper	&	Lower	&		&		&		&		&		&		&		&		&		&		\\
\#	&	level ($k$)	&	level ($i$)	&	$\lambda_\mathrm{TW}$	&	$\lambda_\mathrm{NIST}$	&	$A_{\mathrm{TW,l}}$	&	$A_{\mathrm{TW,v}}$	&	$A_\mathrm{G}$	&	$A_\mathrm{M}$	&	$A_\mathrm{BH}$	&	$A_\mathrm{Z}$	&	$A_\mathrm{FT}$	&	$A_\mathrm{Exp_\mathrm{Av}}$	\\
\hline																									
	&		&		&		&		&		&		&		&		&		&		&		&		\\
	&	\multicolumn{2}{c}{$(^{3}P)3p-(^{3}P)3s$} 																							\\
1	&	$^{2}S_{1/2}$	&	$^{2}P_{1/2}$	&	356.57	&	355.78	&	0.29	&	0.27	&	0.75	&	0.22	&	0.44	&	$\cdots$	&	0.22	&	0.22$\pm$0.01	\\
2	&		&	$^{2}P_{3/2}$	&	349.62	&	348.19	&	1.53	&	1.31	&	0.77	&	1.44	&	1.19	&	1.13	&	1.39	&	1.45$\pm$0.08	\\
3	&	$^{2}P_{1/2}$	&	$^{2}P_{1/2}$	&	325.46	&	337.82	&	1.54	&	1.70	&	0.85	&	1.52	&	1.19	&	1.25	&	1.48	&	1.49$\pm$0.09	\\
4	&		&	$^{2}P_{3/2}$	&	324.75	&	330.97	&	0.50	&	0.55	&	0.94	&	0.22	&	0.51	&	$\cdots$	&	0.26	&	0.24$\pm$0.02	\\
5	&	$^{2}P_{3/2}$	&	$^{2}P_{1/2}$	&	326.39	&	339.28	&	0.44	&	0.47	&	0.40	&	0.39	&	0.32	&	0.31	&	0.39	&	0.38$\pm$0.02	\\
6	&		&	$^{2}P_{3/2}$	&	325.68	&	332.37	&	1.63	&	1.81	&	1.35	&	1.42	&	1.42	&	1.62	&	1.38	&	1.40$\pm$0.08	\\
7	&	$^{2}D_{3/2}$	&	$^{2}P_{1/2}$	&	372.43	&	372.71	&	1.18	&	1.12	&	0.98	&	1.07	&	1.10	&	1.16	&	1.06	&	0.93$\pm$0.05	\\
8	&		&	$^{2}P_{3/2}$	&	365.51	&	364.39	&	0.34	&	0.30	&	0.33	&	0.33	&	0.30	&	0.25	&	0.33	&	0.36$\pm$0.02	\\
9	&	$^{2}D_{5/2}$	&	$^{2}P_{3/2}$	&	384.63	&	371.31	&	1.30	&	1.32	&	1.29	&	1.38	&	1.41	&	1.40	&	1.37	&	1.14$\pm$0.06	\\
10	&	$^{4}S_{3/2}$	&	$^{4}P_{1/2}$	&	306.92	&	302.89	&	0.52	&	0.46	&	0.32	&	0.51	&	0.55	&	0.40	&	0.45	&	0.43$\pm$0.03	\\
11	&		&	$^{4}P_{3/2}$	&	304.66	&	300.17	&	0.91	&	0.78	&	0.76	&	0.92	&	0.90	&	0.81	&	0.84	&	0.75$\pm$0.05	\\
12	&		&	$^{4}P_{5/2}$	&	299.21	&	295.57	&	1.16	&	0.96	&	1.40	&	1.25	&	1.13	&	1.27	&	1.12	&	0.96$\pm$0.06	\\
13	&	$^{4}P_{1/2}$	&	$^{4}P_{1/2}$	&	363.97	&	375.12	&	0.20	&	0.23	&	0.18	&	0.19	&	0.19	&	0.21	&	0.18	&	0.17$\pm$0.01	\\
14	&		&	$^{4}P_{3/2}$	&	377.62	&	370.96	&	1.09	&	1.22	&	1.10	&	1.20	&	1.23	&	1.07	&	1.14	&	1.04$\pm$0.05	\\
15	&	$^{4}P_{3/2}$	&	$^{4}P_{1/2}$	&	365.99	&	377.71	&	0.47	&	0.53	&	0.47	&	0.44	&	0.86	&	0.51	&	0.42	&	0.35$\pm$0.02	\\
16	&		&	$^{4}P_{3/2}$	&	362.79	&	373.49	&	0.23	&	0.25	&	0.20	&	0.20	&	0.19	&	0.17	&	0.19	&	0.16$\pm$0.01	\\
17	&		&	$^{4}P_{3/2}$	&	359.94	&	366.41	&	0.81	&	0.82	&	0.45	&	0.75	&	0.80	&	0.60	&	0.71	&	0.62$\pm$0.03	\\
18	&	$^{4}P_{5/2}$	&	$^{4}P_{3/2}$	&	365.98	&	376.63	&	0.31	&	0.35	&	0.29	&	0.31	&	0.31	&	0.37	&	0.30	&	0.27$\pm$0.01	\\
19	&		&	$^{4}P_{5/2}$	&	375.24	&	369.42	&	1.01	&	1.11	&	1.02	&	1.08	&	1.11	&	0.90	&	1.03	&	0.90$\pm$0.04	\\
20	&	$^{4}D_{1/2}$	&	$^{4}P_{1/2}$	&	339.23	&	334.44	&	1.61	&	1.39	&	1.54	&	1.64	&	1.67	&	1.48	&	1.53	&	1.41$\pm$0.21	\\
21	&		&	$^{4}P_{3/2}$	&	334.31	&	331.13	&	0.23	&	0.22	&	$\cdots$	&	0.27	&	0.28	&	0.30	&	0.26	&	0.27$\pm$0.05	\\
22	&	$^{4}D_{3/2}$	&	$^{4}P_{1/2}$	&	336.20	&	336.06	&	0.91	&	0.84	&	0.84	&	0.87	&	0.88	&	0.73	&	0.82	&	0.84$\pm$0.07	\\
23	&		&	$^{4}P_{3/2}$	&	333.49	&	332.72	&	0.94	&	0.84	&	0.89	&	0.97	&	1.00	&	0.96	&	0.91	&	0.90$\pm$0.07	\\
24	&		&	$^{4}P_{5/2}$	&	327.68	&	327.08	&	0.05	&	0.04	&	0.06	&	0.06	&	0.06	&	0.09	&	0.06	&	0.05$\pm$0.008	\\
25	&	$^{4}D_{5/2}$	&	$^{2}P_{3/2}$	&	389.36	&	394.23	&	0.01	&	0.01	&	0.01	&	0.01	&	0.003	&	$\cdots$	&	0.01	&	0.01$\pm$0.002	\\
26	&		&	$^{4}P_{3/2}$	&	335.67	&	335.50	&	1.47	&	1.35	&	1.35	&	1.43	&	1.47	&	1.23	&	1.34	&	1.27$\pm$0.19	\\
27	&		&	$^{4}P_{5/2}$	&	329.78	&	329.77	&	0.42	&	0.37	&	0.42	&	0.46	&	0.46	&	0.55	&	0.47	&	0.42$\pm$0.06	\\
28	&	$^{4}D_{7/2}$	&	$^{4}P_{5/2}$	&	334.23	&	333.48	&	1.88	&	1.81	&	1.81	&	1.92	&	1.95	&	1.77	&	1.80	&	1.77$\pm$0.15	\\
	&		&		&		&		&		&		&		&		&		&		&		&		\\
	&	\multicolumn{2}{c}{ $(^{1}D)3p-(^{1}D)3s$}																							\\
29	&	$^{2}P_{3/2}$	&	$^{2}D_{3/2}$	&	324.22	&	334.58	&	0.23	&	0.27	&	0.17	&	0.25	&	0.05	&	$\cdots$	&	0.23	&	0.18$\pm$0.04	\\
30	&		&	$^{2}D_{5/2}$	&	324.18	&	334.55	&	1.79	&	2.17	&	1.50	&	1.53	&	1.75	&	$\cdots$	&	1.47	&	1.33$\pm$0.28	\\
31	&	$^{2}F_{5/2}$	&	$^{2}D_{3/2}$	&	356.97	&	357.46	&	1.49	&	1.42	&	1.30	&	1.44	&	1.51	&	$\cdots$	&	1.40	&	1.23$\pm$0.08	\\
32	&		&	$^{2}D_{5/2}$	&	356.94	&	357.42	&	0.13	&	0.11	&	0.09	&	0.11	&	0.11	&	$\cdots$	&	0.11	&	0.10$\pm$0.01	\\
33	&	$^{2}F_{7/2}$	&	$^{2}D_{5/2}$	&	350.86	&	356.85	&	1.68	&	1.53	&	1.30	&	1.57	&	1.63	&	$\cdots$	&	1.51	&	1.29$\pm$0.09	\\
\hline																									
\hline																									
\end{tabular}																									
%
%
\end{table*}																									
%
%
%
%
\begin{table*}																													
%
%
  \caption{Wavelengths $\lambda$ (in nm) and transition probabilities
    $A_{ki}$ (in 10$^{8}$ s$^{-1}$) for some $3d-3p$ lines and
    comparisons with other theoretical and experimental results.					
    $\lambda_\mathrm{TW}$, $A_{\mathrm{TW,l}}$ and
    $A_{\mathrm{TW,v}}$ represent, respectively, the
    wavelength and the transition probability values in
    the length and velocity gauges calculated in this
    work.  																										
    $A_\mathrm{G}$, $A_\mathrm{M}$, $A_\mathrm{BH}$,
    $A_\mathrm{FH}$, $A_\mathrm{GF}$, $A_\mathrm{Z}$,
    $A_\mathrm{FT}$ denote theoretical values taken from			
    Garstang~\cite{1227}, Marantz~\cite{M007}, Blackford and
    Hibbert~\cite{1200}, Fischer and He~\cite{1027}, Godefroid and
    Fischer~\cite{1036}, Zheng~\cite{1026},  and Fischer and
    Tachiev~\cite{1107}, respectively.
    $\lambda_\mathrm{NIST}$ represent the experimental NIST values
    by Kramida and Nave~\cite{1647}, and
    $A_\mathrm{Exp_\mathrm{Av}}$  the weighted average
    experimental values.
    \protect\label{tab_02}} 
%
\begin{tabular}{cccccccccccccccc}																													
\hline																													
\hline																													
	&	Upper	&	Lower																									\\
\#	&	level ($k$)	&	level ($i$)	&	$\lambda_\mathrm{TW}$	&	$\lambda_\mathrm{NIST}$	&	$A_\mathrm{TW,l}$	&	$A_\mathrm{TW,v}$	&	$A_\mathrm{G}$	&	$A_\mathrm{M}$	&	$A_\mathrm{BH}$	&	$A_\mathrm{FH}$	&	$A_\mathrm{GF}$	&	$A_\mathrm{Z}$	&	$A_\mathrm{FT}$	&	$A_\mathrm{Exp_\mathrm{Av}}$	\\
\hline																													
	&		&		&		&		&		&		&		&		&		&		&		&		&		&		\\
	&	\multicolumn{2}{c}{$(^{3}P)3d-(^{3}P)3p$} 																											\\
1	&	$^{2}P_{1/2}$	&	$^{2}S_{1/2}$	&	357.68	&	350.36	&	2.01	&	2.12	&	1.90	&	1.99	&	2.01	&	2.03	&	$\cdots$	&	1.65	&	2.10	&	2.32$\pm$0.14	\\
2	&	$^{2}D_{3/2}$	&	$^{2}P_{1/2}$	&	397.81	&	381.84	&	0.71	&	0.68	&	0.69	&	0.64	&	1.06	&	0.67	&	$\cdots$	&	$\cdots$	&	0.68	&	0.56$\pm$0.05	\\
3	&		&	$^{2}P_{3/2}$	&	366.58	&	380.00	&	0.14	&	0.11	&	$\cdots$	&	0.36	&	0.57	&	0.35	&	$\cdots$	&	0.30	&	0.37	&	0.30$\pm$0.03	\\
4	&	$^{2}D_{5/2}$	&	$^{2}P_{3/2}$	&	399.01	&	382.98	&	0.84	&	0.73	&	0.88	&	0.93	&	1.68	&	0.90	&	$\cdots$	&	$\cdots$	&	0.98	&	0.79$\pm$0.07	\\
5	&		&	$^{2}D_{3/2}$	&	355.28	&	347.76	&	0.27	&	0.30	&	0.34	&	0.44	&	0.50	&	0.33	&	$\cdots$	&	$\cdots$	&	0.33	&	0.30$\pm$0.03	\\
6	&		&	$^{4}D_{3/2}$	&	337.51	&	326.99	&	0.29	&	0.29	&	0.48	&	0.24	&	0.00	&	0.39	&	$\cdots$	&	$\cdots$	&	0.37	&	0.3$\pm$0.09	\\
7	&	$^{2}F_{5/2}$	&	$^{2}P_{3/2}$	&	391.94	&	375.38	&	0.21	&	0.18	&	0.55	&	0.52	&	0.12	&	0.05	&	$\cdots$	&	$\cdots$	&	0.16	&	0.26$\pm$0.06	\\
8	&		&	$^{4}D_{3/2}$	&	307.34	&	321.43	&	0.01	&	0.01	&	1.61	&	0.65	&	0.00	&	0.14	&	$\cdots$	&	$\cdots$	&	0.20	&	0.75$\pm$0.23	\\
9	&	$^{2}F_{7/2}$	&	$^{2}D_{5/2}$	&	357.05	&	340.69	&	2.00	&	2.10	&	1.10	&	1.47	&	3.61	&	1.64	&	1.89	&	$\cdots$	&	1.61	&	1.16$\pm$0.06	\\
10	&		&	$^{4}D_{7/2}$	&	311.85	&	320.90	&	0.10	&	0.09	&	0.12	&	0.15	&	0.00	&	0.16	&	0.19	&	$\cdots$	&	0.16	&	0.11$\pm$0.01	\\
11	&	$^{4}P_{1/2}$	&	$^{4}P_{1/2}$	&	293.32	&	292.56	&	0.56	&	0.48	&	0.52	&	0.56	&	0.56	&	0.41	&	$\cdots$	&	$\cdots$	&	0.49	&	0.52$\pm$0.16	\\
12	&		&	$^{4}P_{3/2}$	&	291.01	&	298.40	&	1.57	&	1.38	&	$\cdots$	&	$\cdots$	&	0.49	&	1.11	&	$\cdots$	&	1.48	&	1.46	&	1.70$\pm$0.51	\\
13	&		&	$^{4}S_{3/2}$	&	373.65	&	379.55	&	1.28	&	1.42	&	$\cdots$	&	$\cdots$	&	1.54	&	1.16	&	$\cdots$	&	1.54	&	1.47	&	1.30$\pm$0.39	\\
14	&	$^{4}P_{3/2}$	&	$^{4}S_{3/2}$	&	356.58	&	369.51	&	0.96	&	1.05	&	0.82	&	1.06	&	1.67	&	0.30	&	$\cdots$	&	$\cdots$	&	1.01	&	0.57$\pm$0.04	\\
15	&		&	$^{4}P_{5/2}$	&	302.96	&	287.30	&	0.48	&	0.43	&	0.46	&	0.63	&	0.96	&	0.08	&	$\cdots$	&	$\cdots$	&	0.45	&	0.34$\pm$0.10	\\
16	&		&	$^{4}D_{1/2}$	&	308.05	&	320.94	&	0.11	&	0.11	&	0.51	&	0.17	&	0.05	&	1.48	&	$\cdots$	&	$\cdots$	&	0.34	&	0.39$\pm$0.12	\\
17	&		&	$^{4}D_{3/2}$	&	319.46	&	334.05	&	0.29	&	0.30	&	0.14	&	0.34	&	0.11	&	0.96	&	$\cdots$	&	$\cdots$	&	0.49	&	0.29$\pm$0.09	\\
18	&	$^{4}P_{5/2}$	&	$^{4}S_{3/2}$	&	365.35	&	354.28	&	1.32	&	1.41	&	1.30	&	1.33	&	1.94	&	1.22	&	$\cdots$	&	$\cdots$	&	1.49	&	1.29$\pm$0.10	\\
19	&		&	$^{4}P_{3/2}$	&	293.82	&	287.63	&	0.64	&	0.58	&	0.84	&	0.83	&	0.73	&	0.18	&	$\cdots$	&	$\cdots$	&	0.70	&	0.22$\pm$0.06	\\
20	&		&	$^{4}P_{5/2}$	&	300.37	&	285.80	&	0.72	&	0.67	&	0.91	&	0.84	&	1.23	&	0.05	&	$\cdots$	&	$\cdots$	&	0.72	&	0.25$\pm$0.07	\\
21	&		&	$^{2}D_{3/2}$	&	348.54	&	337.18	&	0.25	&	0.25	&	0.12	&	0.18	&	0.00	&	0.71	&	$\cdots$	&	$\cdots$	&	0.17	&	0.23$\pm$0.07	\\
22	&		&	$^{4}D_{3/2}$	&	313.05	&	317.61	&	0.03	&	0.25	&	0.03	&	0.15	&	0.04	&	0.14	&	$\cdots$	&	$\cdots$	&	0.06	&	0.06$\pm$0.02	\\
23	&	$^{4}D_{1/2}$	&	$^{4}P_{1/2}$	&	308.65	&	304.56	&	2.40	&	0.02	&	2.50	&	2.56	&	3.00	&	2.50	&	$\cdots$	&	2.37	&	2.40	&	2.77$\pm$0.59	\\
24	&		&	$^{4}P_{3/2}$	&	319.84	&	302.87	&	0.73	&	0.72	&	0.84	&	0.85	&	0.88	&	0.82	&	$\cdots$	&	$\cdots$	&	0.80	&	0.60$\pm$0.18	\\
25	&		&	$^{4}D_{3/2}$	&	336.29	&	347.13	&	0.32	&	0.35	&	$\cdots$	&	$\cdots$	&	0.41	&	0.39	&	$\cdots$	&	0.39	&	0.37	&	0.35$\pm$0.11	\\
26	&		&	$^{4}D_{1/2}$	&	337.93	&	341.09	&	0.26	&	0.29	&	$\cdots$	&	$\cdots$	&	0.36	&	0.29	&	$\cdots$	&	0.38	&	0.30	&	0.30$\pm$0.09	\\
27	&	$^{4}D_{3/2}$	&	$^{4}P_{1/2}$	&	322.26	&	305.47	&	0.80	&	0.82	&	0.93	&	0.98	&	1.21	&	0.96	&	$\cdots$	&	1.17	&	0.92	&	0.87$\pm$0.18	\\
28	&		&	$^{4}P_{3/2}$	&	315.53	&	303.77	&	1.84	&	1.80	&	2.00	&	2.04	&	2.27	&	1.97	&	$\cdots$	&	$\cdots$	&	1.91	&	2.02$\pm$0.19	\\
29	&		&	$^{4}P_{5/2}$	&	313.13	&	301.73	&	0.34	&	0.33	&	0.35	&	0.35	&	0.35	&	0.34	&	$\cdots$	&	$\cdots$	&	0.33	&	0.64$\pm$0.19	\\
30	&		&	$^{4}D_{3/2}$	&	337.41	&	348.16	&	0.27	&	0.30	&	0.38	&	0.31	&	0.37	&	0.29	&	$\cdots$	&	$\cdots$	&	0.30	&	0.27$\pm$0.03	\\
31	&	$^{4}D_{5/2}$	&	$^{4}P_{3/2}$	&	306.41	&	304.76	&	1.68	&	1.58	&	1.80	&	1.82	&	2.18	&	1.82	&	$\cdots$	&	1.98	&	1.74	&	1.87$\pm$0.40	\\
32	&		&	$^{4}P_{5/2}$	&	309.36	&	302.70	&	1.47	&	1.36	&	1.50	&	1.48	&	1.54	&	1.38	&	$\cdots$	&	$\cdots$	&	1.36	&	1.46$\pm$0.14	\\
33	&		&	$^{4}D_{5/2}$	&	335.78	&	346.58	&	0.48	&	0.51	&	0.55	&	0.48	&	0.62	&	0.49	&	$\cdots$	&	0.44	&	0.49	&	0.49$\pm$0.06	\\
34	&	$^{4}D_{7/2}$	&	$^{4}P_{5/2}$	&	303.45	&	320.41	&	2.65	&	2.66	&	3.10	&	3.16	&	3.53	&	3.06	&	3.02	&	2.84	&	2.95	&	2.73$\pm$0.21	\\
35	&		&	$^{4}D_{7/2}$	&	342.25	&	332.92	&	0.87	&	0.93	&	0.87	&	0.89	&	1.07	&	0.92	&	$\cdots$	&	0.67	&	0.89	&	0.87$\pm$0.06	\\
36	&	$^{4}F_{3/2}$	&	$^{4}S_{3/2}$	&	363.03	&	357.12	&	0.03	&	0.04	&	0.43	&	0.27	&	0.00	&	1.30	&	$\cdots$	&	$\cdots$	&	0.46	&	0.92$\pm$0.04	\\
37	&		&	$^{4}D_{3/2}$	&	327.81	&	319.89	&	0.59	&	0.60	&	0.59	&	0.48	&	1.06	&	0.00	&	$\cdots$	&	$\cdots$	&	0.36	&	0.32$\pm$0.10	\\
38	&	$^{4}F_{5/2}$	&	$^{2}D_{3/2}$	&	326.88	&	338.84	&	0.02	&	0.02	&	2.00	&	1.64	&	0.00	&	0.09	&	$\cdots$	&	$\cdots$	&	0.08	&	1.53$\pm$0.33	\\
39	&		&	$^{4}D_{3/2}$	&	331.33	&	321.43	&	1.94	&	2.06	&	0.73	&	0.86	&	3.39	&	2.38	&	$\cdots$	&	$\cdots$	&	2.31	&	0.94$\pm$0.28	\\
40	&	$^{4}F_{7/2}$	&	$^{2}D_{5/2}$	&	354.17	&	336.72	&	1.37	&	1.41	&	1.00	&	1.43	&	0.00	&	1.60	&	1.79	&	$\cdots$	&	1.60	&	1.23$\pm$0.09	\\
41	&		&	$^{4}D_{5/2}$	&	318.50	&	319.86	&	1.69	&	1.40	&	2.30	&	1.87	&	3.86	&	1.66	&	1.49	&	$\cdots$	&	1.61	&	0.98$\pm$0.08	\\
42	&	$^{4}F_{9/2}$	&	$^{4}D_{7/2}$	&	328.89	&	321.82	&	3.52	&	3.58	&	3.60	&	3.70	&	4.23	&	3.73	&	$\cdots$	&	3.45	&	3.65	&	3.66$\pm$0.63	\\
\hline																													
\hline																													
\end{tabular}																													
%
%
\end{table*}																													

The results obtained in this work for the 1s$^{2}$ 2s$^{2}$
  2p$^{4}$ 3p $^{4}$P$_{7/2} - $1s$^{2}$ 2s$^{2}$ 2p$^{4}$ 3s
  transition wavelength, shown in Table~\ref{tab_A}, are an example of
  the importance of the inclusion of correlation. We observe that the
  calculated single-configuration Dirac-Fock wavelength differs by
  about 4\% from the experimental value. The inclusion of correlation
  narrows this difference to less than 0.3\%.
  It is worthwhile to call attention to the fact that the correlation
  configuration 2p$^2$ 3p $6 \ell^2$, in the upper level, and the 2p$^2$
  3s $6 \ell^2$ one in the lower level, with $\ell=$s...f, give only
  a contribution of 0.25 nm to the wavelength, which is 0.08\% of the
  calculated value. Furthermore, we observe that the inclusion of the
  mentioned configurations with $\ell=$f gives no contribution to the
  wavelength value.

  Considering this analysis, 
  to obtain the valence and the core-valence correlation contributions
  we used a virtual space spanned by single and double-excited
  configurations up to 3d orbitals, resulting from the excitation of
  $n=2$ and $n=3$ electrons in the upper and lower levels, and the
  double-excited configurations 2p$^2$ 3p $n^{\prime} \ell^2$, for the upper
  level, and 2p$^2$ 3s $n^{\prime} \ell^2$, with $n^{\prime}
  \ell^2=$4s$^2$...6d$^2$. 
%
  This educated choice allowed us to maintain a manageable virtual
  orbital space, and to avoid the nonrelativistic offset through the
  inclusion of the ``Brillouin'' single excitations~\cite{1757}.  
%
%
%
%

\begin{figure}
  \centering
  \includegraphics[width=\textwidth]{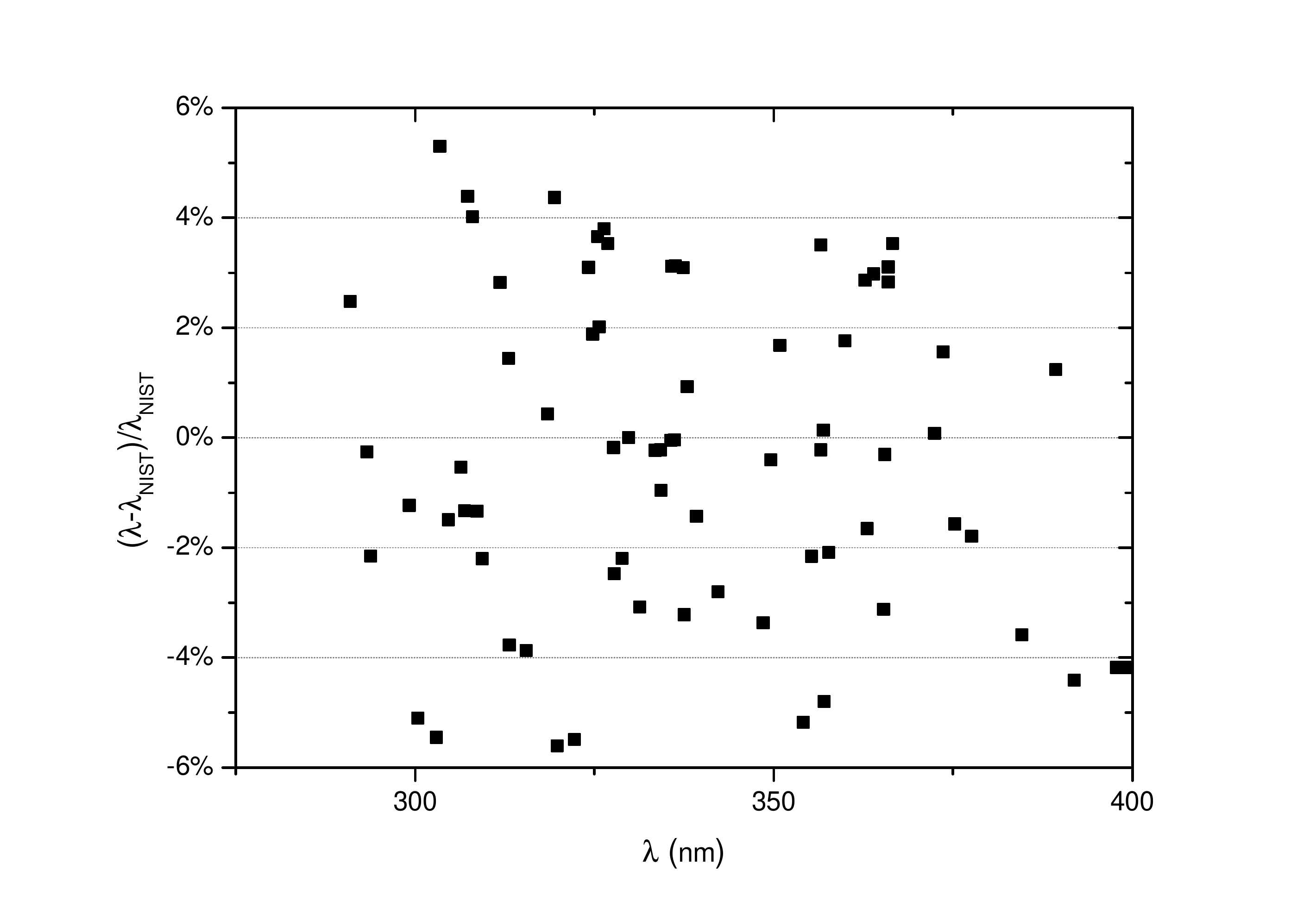}
  \caption{Relative difference between the theoretical wavelengths
  calculated in this work and the experimental results by Kramida and
  Nave~\cite{1647}.
  \label{figure_L}}
\end{figure}

%
A comparison between the results of this work for the transition
wavelengths and Kramida and Nave's experimental results~\cite{1647},
published in the NIST webpage~\cite{853}, is presented in
Tables~\ref{tab_01} and ~\ref{tab_02}, and illustrated in
Figure~\ref{figure_L}. 
We observe that all theoretical results differ by less than 6\% from
the experimental ones, and more than 20\% differ by less than 1\% from
the Kramida and Nave's results.  This agreement validates our
transition wavelength results.

The role of gauge invariance in the interaction between the
electromagnetic field and the electron-positron field is discussed
explicitly in well-known texts~\cite{12}. Existing relativistic
self-consistent-field calculations of radiative atomic transition
probabilities have been carried out in the Coulomb or Babushkin
gauges. 
In the nonrelativistic limit the Coulomb gauge formula for the
transition probability yields the dipole velocity expression whilst
the Babushkin formula gives the dipole length expression~\cite{6}. 
%

From the point of view of the transition probability accuracy, an
important requisite is the agreement between length and velocity
forms. Nevertheless, one should be cautious about drawing conclusions
from such an agreement in the relativistic case, since it depends on
the proper inclusion of the negative energy state ~\cite{737}, which
cannot be explicitly done in the present calculations.
%
%
When there is no good agreement between the two forms, there are
reasons for a preference of the length form over the velocity form
from the non-relativistic ~\cite{13,20} and relativistic~\cite{1471}
points of view.

\begin{figure*}
  \centering
  \includegraphics[width=\textwidth]{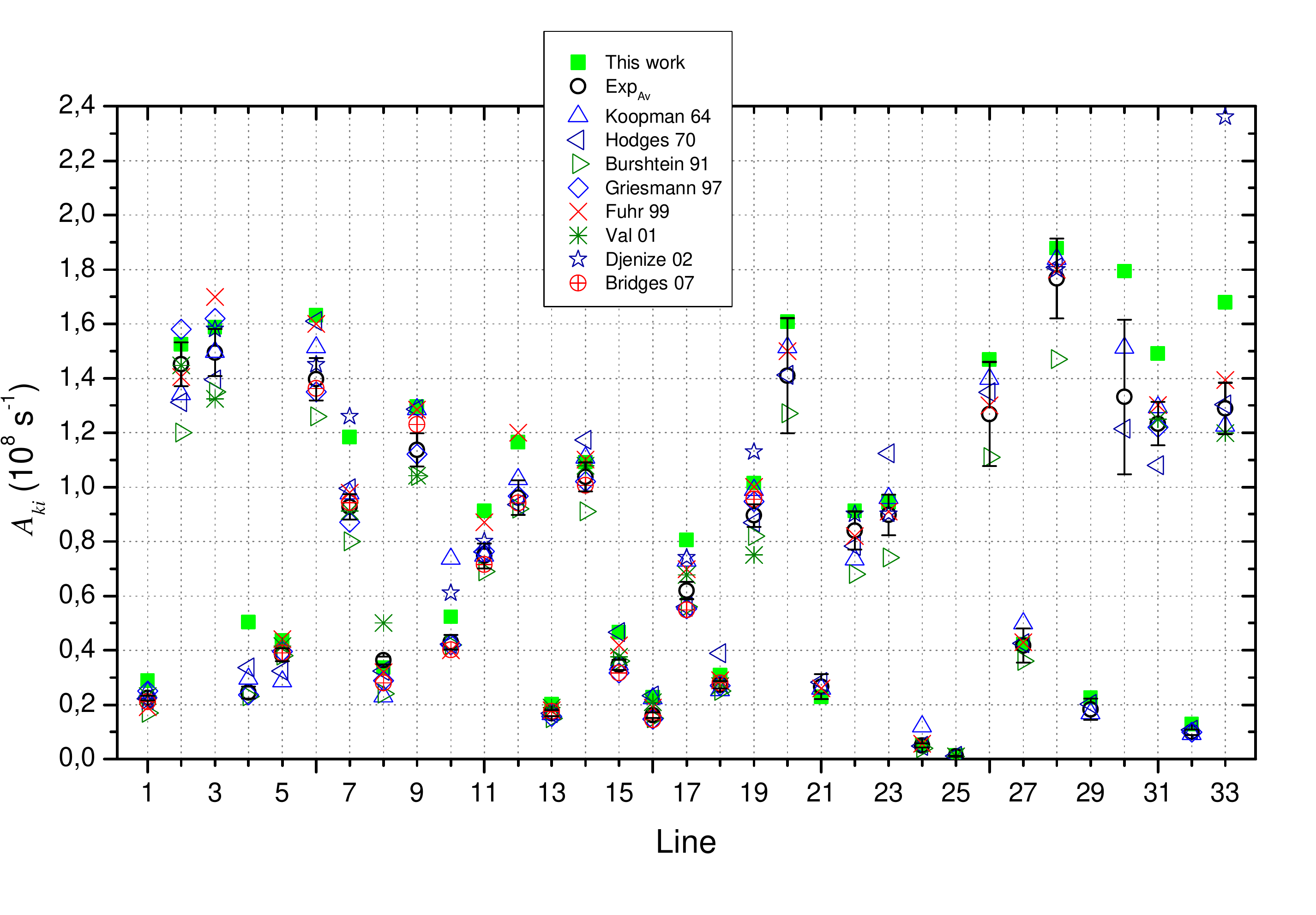}
  \caption{Comparison of our theoretical results and the available
    experimental results for $3p-3s$ transition probability values.
  \label{figure_1}}
\end{figure*}
\begin{figure*}
\centering
\includegraphics[width=\textwidth]{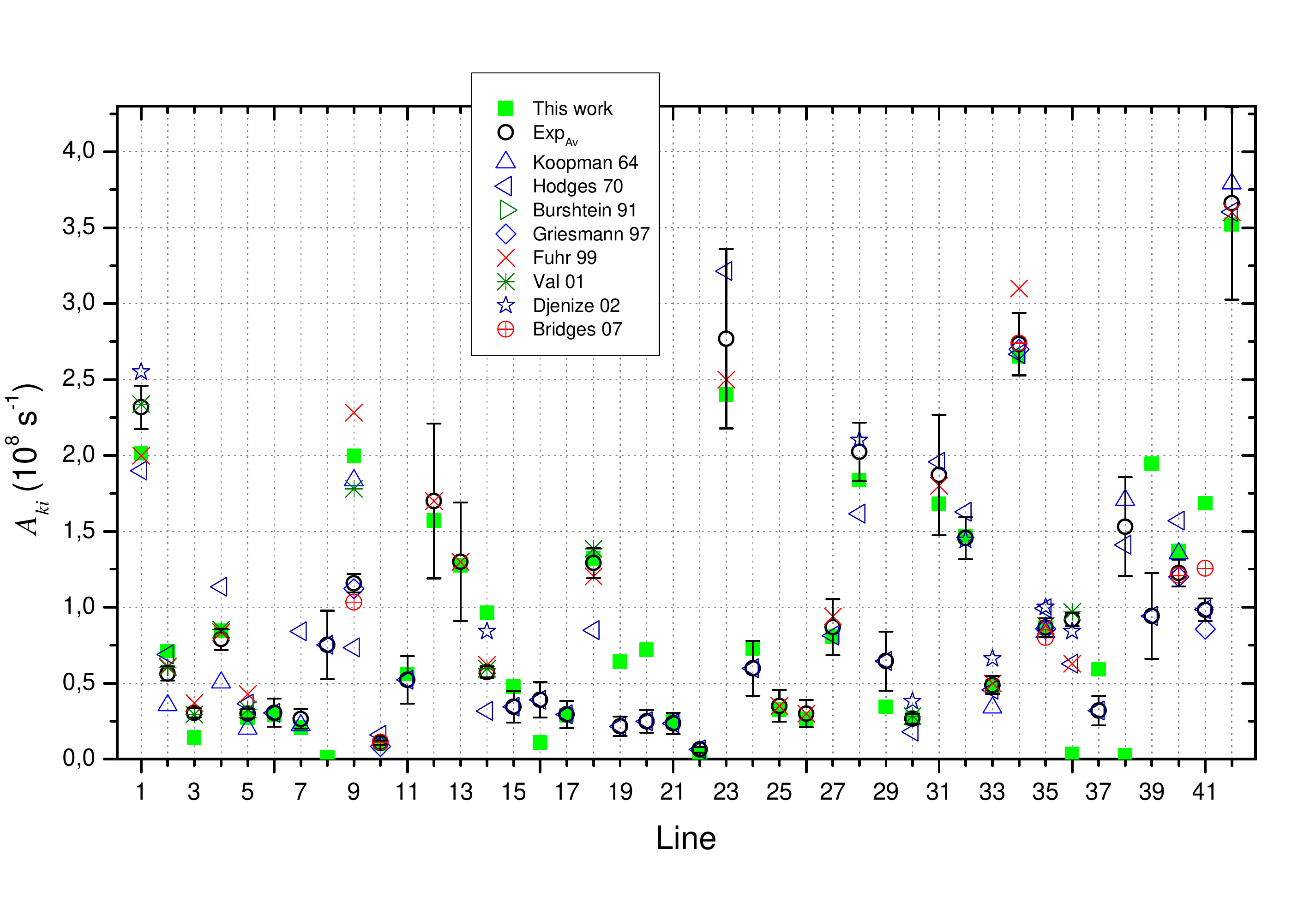}
\caption{Comparison of our theoretical results and the available
  experimental results for $3d-3p$ transition probability values.
  \label{figure_2}}
\end{figure*}

A graphical comparison of our results with 
experimental results for the transition probabilities of the 3p-3s and
3d-3p is shown in Fig. ~\ref{figure_1} and \ref{figure_2},
respectively. The line identification is given in Tables~\ref{tab_01}
and \ref{tab_02}.
The first conclusion we draw from these figures is that there is a large
disagreement between the experimental results. The maximum relative
difference between the experimental values for each transition spans a
range from 45\% to 200\% in the 3p-3s transitions, and from 5\% to
277\% in the 3d-3p transitions.  The average of these maxima is 45\%
for the former transitions and 71\% for the latter ones.
%
The cause of these discrepancies, even within experiments of the same
type is puzzling. Possible reasons are misidentifications of the
lines, normalization problems, and nonselective excitation of atomic
energy levels.

In order to have an experimental reference to assess the theoretical calculations we computed, for each transition, the weighted average, $A_\mathrm{Exp_\mathrm{Av}}$, using as weight, for each case, the inverse of the square of the uncertainty. 

In the cases the uncertainty is not provided, such as Koopman~\cite{1029}, Hodges and Marantz~\cite{1115}, Burshtein and Vujnovic~\cite{1039}, and Fuhr and Wiese~\cite{1113}, we have assumed an uncertainty of 30\%.

In Tables~\ref{tab_01} and ~\ref{tab_02}, the transition probability values ($A_{ki}$) calculated in this work for several ($^3$ P)3p -- ($^3$ P)3s and ($^3$ P)3d -- ($^3$ P)3p lines ($LS$ dipole-allowed and intersystem lines), respectively, in Ne II are compared with the available theoretical values of Garstang~\cite{1227}, Marantz ~\cite{M007}, Blackford and Hibbert~\cite{1200}, Zheng ~\cite{1026}, and Fischer and Tachiev~\cite{1107}, and with the experimental weighted average (EWA), $A_\mathrm{Exp_\mathrm{Av}}$.

A general agreement between length and velocity forms of the transition probabilities is found; in 91\% of our results the length and velocity forms of the oscillator strengths differ by less than 20\% and in 60\% of our results the two forms differ by less than10\%.

Comparing our ab initio values and the adjusted results by Fischer and Tachiev~\cite{1107} to the EWA values, in 50\% of the  3d-3p lines our ab initio values are closer than the later adjusted values, and in 80\% of the 3p-3s lines the adjusted Fischer and Tachiev values are closer than our values. In order to assess the influence of the inclusion of the experimental transition energies in the adjusted $A_{ki}$ values, we present in Table~\ref{tab_04}, for some lines, the relative differences 
between our ab initio values $A_{\mathrm{TW,l }}$, our adjusted values with the NIST transition energies $A_{\mathrm{TW,l\ adjusted}}$, the adjusted Fischer and Tachiev values, and the weighted average experimental values $A_\mathrm{Exp_\mathrm{Av}}$. 
We observe that in all analyzed cases the energy adjustment bring our ab initio results closer to the EWA values than the Fischer and Tachiev results. Nevertheless, we must emphasize that this agreement is only indicative, and should not be considered in an absolute way since, as mentioned earlier, there is a large disagreement between the experimental results and, consequently, the EWA reflect only an experimental trend.
%


\begin{table}																		
\begin{center}																		
\caption{Relative differences $R_{\mathrm{TW}}$, $R_{\mathrm{TW,adjusted}}$, and $R_{\mathrm{FT}}$ between the theoretical transition probability values $A_{\mathrm{TW,l }}$,  $A_{\mathrm{TW,l\ adjusted}}$, and $A_\mathrm{FT}$, and the $A_\mathrm{Exp_\mathrm{Av}}$ weighted average experimental values, respectively. 																
\protect\label{tab_04}}																		
%
%
\begin{tabular}{cccccccccccccc}																		
\hline																		
\hline																		
	&	Upper	&	Lower	&			&			&			\\
\#	&	level ($k$)	&	level ($i$)	&	$R_{\mathrm{TW}}$		&	$R_{\mathrm{TW,adjusted}}$		&	$R_\mathrm{FT}$		\\
\hline														
	&		&		&			&			&			\\
	&	\multicolumn{2}{c}{$(^{3}P)3p-(^{3}P)3s$} 												\\
5	&	$^{2}P_{3/2}$	&	$^{2}P_{1/2}$	&	-10	\%	&	2	\%	&	-2	\%	\\
18	&	$^{4}P_{5/2}$	&	$^{4}P_{3/2}$	&	-13	\%	&	-3	\%	&	-10	\%	\\
	&		&		&			&			&			\\
	&	\multicolumn{2}{c}{ $(^{1}D)3p-(^{1}D)3s$}												\\
29	&	$^{2}P_{3/2}$	&	$^{2}D_{3/2}$	&	-24	\%	&	-13	\%	&	-26	\%	\\
\hline																		
\hline																		
\end{tabular}																		
\end{center}																		
\end{table}																		

\clearpage 
%
%

\clearpage
%
%
\clearpage
%
%
 
%
%
%
\section{Conclusions}
\label{sec4}

We presented \textit{ab initio} relativistic calculated values of
transition wavelengths and probabilities for several 2p$^4$ 3p -
2p$^4$ 3s and 2p$^4$ 3d - 2p$^4$ 3p lines Ne II.
These transitions are of interest because of their importance in the
interpretation of stellar thermonuclear reactions.
We showed the importance of the inclusion of correlation in a balanced
way.
The results are compared with existing theoretical calculations and
with experimental data. The dispersion of experimental results for
each analyzed transition prevents the assessment of the available
theoretical data, calling for more accurate experimental results.
%
%
%
\section*{Acknowledgments}
  This research was supported in part by FCT 
  (Portugal), by the French-Portuguese Collaboration (PESSOA Hubert Curien Program,
  Contract N$^{\mathrm{o}}$ 10721NF and 20022VB), and by the Ac{\c{c}}{\~{o}}es
  Integradas Luso-Francesas (Contract N$^{\mathrm{o}}$
  F-11/09). Laboratoire Kastler Brossel (LKB) is \textquotedblleft
  Unit\'{e} Mixte de Recherche du CNRS, de l'ENS et de l'UPMC
  N$^{\circ}$ 8552\textquotedblright. The LKB group acknowledges the
  support of the Allianz Program of the Helmholtz Association,
  contract EMMI HA-216 \textquotedblleft Extremes of Density and
  Temperature: Cosmic Matter in the Laboratory\textquotedblright.

%
%
%

\bibliography{jps}

\begin{thebibliography}{10}

\bibitem{3258}
D.~E. Osterbrock, {\em Astrophysics of Gaseous Nebulae and Active Galactic
  Nuclei} (University of Science, Mill Valley, CA, 1989).

\bibitem{3259}
V. Trimble, Astronomy and Astrophysics Review {\bf 3},  1  (1991).

\bibitem{1029}
D.~W. Koopman, J. Opt. Soc. Am. A {\bf 54},  1354  (1964).

\bibitem{133}
E.~R. Condon G.~H. Shortley, {\em The Theory of Atomic Spectra} (Cambridge
  University Press, Cambridge, 1953).

\bibitem{853}
NIST Atomic Spectra Database v3.0,
  http://physics.nist.gov/PhysRefData/ASD/index.html, 2008.

\bibitem{1115}
D. Hodges, H. Marantz, C.~L. Tang, J. Opt. Soc. Am. A {\bf 60},  192  (1970).

\bibitem{1039}
M.~L. Burshtein V. Vujnovic, Astron. Astrophys. {\bf 247},  252  (1991).

\bibitem{1193}
U. Griesmann, J. Musielok, W.~L. Wiese, J. Opt. Soc. Am. B {\bf 14},  2204
  (1997).

\bibitem{1113}
J.~R. Fuhr W.~L. Wiese, {\em CRC Handbook of Chemistry and Physics} (CRC Press,
  Boca Raton, Florida, 1999), pp.\ 10--88.

\bibitem{1897}
J.~A. del Val, J.~A. Aparicio, V.~R. Gonzalez, S. Mar, J. Phys. B {\bf 34},
  2513  (2001).

\bibitem{2189}
S. Djenize, V. Milosavljevic, M.~S. Dimitrijevic, Astron. Astrophys. {\bf 382},
   359  (2002).

\bibitem{2122}
J.~M. Bridges W.~L. Wiese, Phys. Rev. A {\bf 76},  022513  (2007).

\bibitem{1227}
R.~H. Garstang, Mon. Not. R. Astron. Soc. {\bf 114},  118  (1954).

\bibitem{M007}
H. Marantz, Ph.D. thesis, Cornell University, 1968, published in ~\cite{1115}.

\bibitem{3351}
F. Herman S. Skillman, {\em Atomic Structure Calculations} (Prentice Hall, USA,
  1963).

\bibitem{1200}
H.~M.~S. Blackford A. Hibbert, At. Data Nucl. Data Tables {\bf 58},  101
  (1994).

\bibitem{1027}
C. Froese~Fischer X. He, Can. J. Phys. {\bf 77},  177  (1999).

\bibitem{1036}
M.~R. Godefroid A. Hibbert, Mol. Phys. {\bf 98},  1099  (2000).

\bibitem{1026}
N.~W. Zheng T. Wang, Spectrochim. Acta Part B {\bf 58},  1319  (2003).

\bibitem{1107}
C. Froese~Fischer G. Tachiev, At. Data Nucl. Data Tables {\bf 87},  1  (2004).

\bibitem{32}
J.~P. Desclaux, Comp. Phys. Commun. {\bf 9},  31  (1975).

\bibitem{62}
P. Indelicato, Phys. Rev. Lett. {\bf 77},  3323  (1996).

\bibitem{3226}
P. Indelicato J. Desclaux, MCDFGME, a MultiConfiguration Dirac Fock and General
  Matrix Elements program (release 2005),
  \url{http://dirac.spectro.jussieu.fr/mcdf}, 2007.

\bibitem{47}
P. Indelicato, Phys. Rev. A {\bf 51},  1132  (1995).

\bibitem{231}
G.~E. Brown D.~E. Ravenhall, Proc. R. Soc. London, Ser. A {\bf 208},  552
  (1951).

\bibitem{229}
J. Sucher, Phys. Rev. A {\bf 22},  348  (1980).

\bibitem{230}
M.~H. Mittleman, Phys. Rev. A {\bf 24},  1167  (1981).

\bibitem{63}
O. Gorceix P. Indelicato, Phys. Rev. A {\bf 37},  1087  (1988).

\bibitem{3275}
G. Audi, A.~H. Wapstra, C. Thibault, Nucl. Phys. A {\bf 729},  337  (2003).

\bibitem{1109}
I. Angeli, At. Data Nucl. Data Tables {\bf 87},  185  (2004).

\bibitem{114}
P.~J. Mohr Y.-K. Kim, Phys. Rev. A {\bf 45},  2727  (1992).

\bibitem{116}
P.~J. Mohr, Phys. Rev. A {\bf 46},  4421  (1992).

\bibitem{3141}
E.-O. Le~Bigot, P. Indelicato, P.~J. Mohr, Phys. Rev. A {\bf 64},  052508
  (2001).

\bibitem{PI_01}
P. Indelicato P. Mohr, Phys. Rev. A {\bf 46},  172  (1992).

\bibitem{117}
P.~J. Mohr G. Soff, Phys. Rev. Lett. {\bf 70},  158  (1993).

\bibitem{58}
P. Indelicato, O. Gorceix, J.~P. Desclaux, Journal of Physics B: Atomic and
  Molecular Physics {\bf 20},  651  (1987).

\bibitem{56}
P. Indelicato J.~P. Desclaux, Phys. Rev. A {\bf 42},  5139  (1990).

\bibitem{53}
P. Indelicato E. Lindroth, Phys. Rev. A {\bf 46},  2426  (1992).

\bibitem{847}
P. Indelicato, S. Boucard, E. Lindroth, Eur. Phys. J. D {\bf 3},  29  (1998).

\bibitem{PI_02}
P. Indelicato P. Mohr, Hyp. Int. {\bf 114},  147  (1998).

\bibitem{782}
P. Indelicato, Hyp. Int. {\bf 108},  39  (1997).

\bibitem{608}
P.-O. Lowdin, Phys. Rev. {\bf 97},  1474  (1955).

\bibitem{1647}
A.~E. Kramida G. Nave, Eur. Phys. J. D {\bf 39},  331  (2006).

\bibitem{1757}
P. Indelicato, E. Lindroth, J.~P. Desclaux, Phys. Rev. Lett. {\bf 94},  013002
  (2005).

\bibitem{12}
A.~I. Akhiezer V.~B. Berestetskii, {\em Quantum Electrodynamics}, {\em
  Interscience Monographs and Texts in Physics and Astronomy} (Interscience
  Publishers, New York, 1965).

\bibitem{6}
I.~P. Grant, Journal of Physics B: Atomic and Molecular Physics {\bf 7},  1458
  (1974).

\bibitem{737}
U.~I. Safronova, W.~R. Johnson, A.~E. Livingston, Phys. Rev. A {\bf 60},  996
  (1999).

\bibitem{13}
A.~F. Starace, Phys. Rev. A {\bf 3},  1242  (1971).

\bibitem{20}
A.~F. Starace, Phys. Rev. A {\bf 8},  1141  (1971).

\bibitem{1471}
A. Derevianko, I.~M. Savukov, W.~R. Johnson, D.~R. Plante, Phys. Rev. A {\bf
  58},  4453  (1998).

\end{thebibliography}

\end{document}